\documentstyle[12pt]{article}
\input psfig.sty
\begin{document}

\title{On the Determination of Proper Time}

\author{Bing Hurl\thanks{Electronic address: heb@alpha02.ihep.ac.cn}\\
{\scriptsize Institute of High Energy Physics, P. O. Box 918(4), 
Beijing 100039, China}\\
\\
Zhiyong Wang\\
{\scriptsize Branch Fact. 1, P.O.Box 1, XinDu, ChenDu, Sichuan 610500,
China}\\
\\
Haidong Zhang\thanks{Electronic address: zhanghd@hptc5.ihep.ac.cn}\\
{\scriptsize Institute of High Energy Physics, P. O. Box 918-4,
Beijing 100039, China}}
\maketitle

\begin{abstract}
Through the analysis of the definition of the duration of proper time 
of a particle given by the length
of its world line, we show that there is no transitivity of
the coordinate time function derived from the definition, so there exists
an ambiguity in the determination of the duration of the proper time for the
particle. Its physical consequence is illustrated with quantum measurement
effect.
\vskip 6mm

\end{abstract}
\newpage
The proper time of a free particle
is referred
to the reading of the clock attached on it.
For a particle moving on spacetime manifold $(M,g_{\mu\nu})$, where the
world line elment $ds$ is given as\footnote[1]{In this paper we always assume
that the temporal and spatial parts of world line element are separable.}
\begin{eqnarray}
ds^2=g_{00}(x)(dx^0)^2-g_{ij}(x)(dx^idx^j),
\end{eqnarray}
for $i,j=1,2,3$,
the duration of its proper time is defined by the length of the world line
element
\begin{eqnarray}
\tau=\int_{\tau_0}^{\tau}ds=\int_{t0}^{t}(h_{00}(x^0)-
h_{ij}(x^0)\frac{dx^i}{dx^0}\frac{dx^j}{dx^0})^{1\over 2}dx^0,
\end{eqnarray}
where $h_{00}(x^0)=g_{00}(x^0,x^i(x^0)), h_{ij}(x^0)=g_{ij}(x^0,x^i(x^0))$,
and $x^i=x^i(x^0)$ is the trajectory of the particle observed in the
coordinate system $\Sigma:(x^0,x^1,x^2,x^3)$. Eq.(2) indicates
that the measure of the duration of proper time should be independent of the
choice of coordinate system and, therefore, it also leads to the relation of
the coordinate times between any a pair of coordinate system $\Sigma$ and
$\Sigma'$:
$$\int_{t_0}^{t}(h_{00}(x_0)-h_{ij}(x^0)\frac{dx^i}{dx^0}\frac{dx^j}{dx^0})
^{1\over 2}dx^0\\$$
\begin{eqnarray}
=\int_{t'_0}^{t'}(h_{0'0'}\left((x^0)'\right)-h_{i'j'}
\left((x^0)'\right)
\frac{(dx^{i})'}{(dx^{0})'}\frac{(dx^{j})'}{(dx^{0})'})^{1\over 2}
(dx^{0})',
\end{eqnarray}
which defines a coordinate time function $t=t(t')$.

For consistency the above-mentioned coordinate time function should be
transitive, i.e. if there are unique $t'(t)$, $t''(t')$ and $t''(t)$ between 
any pair of three systems, we will have $t''\left(t'(t)\right)=t''(t)$.
However, we will prove in this letter that the relation of proper times given by 
Eq.(3) is not transitive, so the matter is much more complicate than we
previously assumed, when we define the the duration of proper time, an
invariant quantity for all observors.

First we prove the existence of coordinate time function. Eq.(3) is an
implicit function of the form: $F(t,t')=0$. The range $(t_0,t)$ is divided
into the union of infinitely many small ones $[t_0,t_1)\cup (t_1,t_2)\cdots
\cup (t_{n-1},t]$. Implicit function theorem gives
a unique function $t'=f_i(t)$ in a small neighborhood $(t_i,t_{i+2})$,
as long as 
\begin{eqnarray}
\frac{\partial F(t,t')}{\partial t'}=(h_{0'0'}\left((x^0)'\right)-h_{i'j'}
\left((x^0)'\right)
\frac{(dx^{i})'}{(dx^{0})'}\frac{(dx^{j})'}{(dx^{0})'})^{1\over 2}
\neq 0.
\end{eqnarray}
By the generalized gluing lemma proved in Appendix we immediately 
obtain a unique continuous time function $t'=t'(t)$
in the range  $(t_0,t)$ and, moreover, it
satisfies 
\begin{eqnarray}
\frac{dt'}{dt}=(h_{00}(t)-h_{ij}
(t)
{dx^{i}\over dt}{dx^{j}\over dt})^{1\over 2}/
(h_{0'0'}(t')-h_{i'j'}(t')
{(dx^{i})'\over dt'} {(dx^{j})'\over dt'})^{1\over 2}
\end{eqnarray}
at every point of the range $[t_0,t]\times [t'_0,t']$.

Suppose there are three different systems $\Sigma^1, \Sigma^2$, and $\Sigma^3$
used to observe the motion of a particle. Without the loss of generality
$\sigma^1$ is the system set up by the co-moving observor of the particle. 
From the previous results
we can obtain the coordinate time relation between $\Sigma^1, \Sigma^2$ as
\begin{eqnarray}
\int_{t_0}^{t}(h^{(1)}_{00}(x^0_{1}))^{1\over 2}dx^0_{1}
=\int_{t'_0}^{t'}(h^{(2)}_{00}(x^0_{2})-h_{ij}^{(2)}(x^0_{2})
\frac{dx^i_{2}}{dx^0_{2}}\frac{dx^j_{2}}{dx^0_{2}})^{1\over 2}
dx^0_{(2)},
\end{eqnarray}
and that between $\Sigma^1,\Sigma^3$ as
\begin{eqnarray}
\int_{t_0}^{t}(h^{(1)}_{00}(x^0_{1}))^{1\over 2}dx^0_{1}
=\int_{t''_0}^{t''}(h^{(3)}_{00}(x^0_{3})-h_{ij}^{(3)}(x^0_{3})
\frac{dx^i_{3}}{dx^0_{3}}\frac{dx^j_{3}}{dx^0_{3}})^{1\over 2}
dx^0_{3}.
\end{eqnarray}

On the other hand, there is the relation of the world line element between
the co-moving system of $\Sigma^2$ and the system $\Sigma^3$:
\begin{eqnarray}
g^{(2)}_{00}(x_{2})(dx_{2}^0)^2=g^{(3)}_{00}(x_{3})(dx^0_{3})^2-
g^{(3)}_{ij}(x_{3})dx^i_{3}dx^j_{3}.
\end{eqnarray}
Substituting Eq.(8) into Eq.(6), we obtain the following coordinate time
relation:
$$\int_{t_0}^{t}(h^{(1)}_{00}(x^0_{1}))^{1\over 2}dx^0_{1}=
\int_{s'_0}^{s'}(g^{(3)}_{00}(x_{3})(dx^0_{3})^2-
g^{(3)}_{ij}(x_{3})dx^i_{3}dx^j_{3}
-g^{(2)}_{ij}(x_{2})dx^i_{2}dx^j_{2})^{1\over 2}$$
\begin{eqnarray}
=\int_{t''_0}^{t''}(h^{(3)}_{00}(x^0_{3})-h_{ij}^{(3)}(x^0_{3})
\frac{dx^i_{3}}{dx^0_{3}}\frac{dx^j_{3}}{dx^0_{3}}-
h_{ij}^{(2)}(x^0_{2}(x^0_{3}))\frac{dx^i_{2}}{dx^0_{2}}
\frac{dx^j_{2}}{dx^0_{2}}
\frac{dx^0_{2}}{dx^0_{3}}\frac{dx^0_{2}}{dx^0_{3}})^{1\over2}
dx^0_{3}.
\end{eqnarray}
With the relation in Eq.(5) between $\Sigma^2$ and $\Sigma^3$, we finally
get another implicit coordinate time 
function:
\begin{eqnarray}
\int_{t_0}^{t}(h^{(1)}_{00}(x^0_{1}))^{1\over 2}dx^0_{1}
=\int_{t''_0}^{t''}(A(x^0_{3})(h^{(3)}_{00}(x^0_{3})-h_{ij}^{(3)}(x^0_{3})
\frac{dx^i_{3}}{dx^0_{3}}\frac{dx^j_{3}}{dx^0_{3}}))^{1\over 2}~dx_{3}^0,
\end{eqnarray}
where $A(x^0_3)=(h_{00}^{(2)}(x^0_{2}(x^0_{3})))
/(h^{(2)}_{00}(x^0_{2}(x^0_{3}))-h_{ij}^{(2)}(x^0_{2}(x^0_{3}))
\frac{dx^i_{2}}{dx^0_{2}}(x^0_2(x^0_3))\frac{dx^j_{2}}{dx^0_{2}}(x^0_2(x^0_3)))
\neq 1$,
between $\Sigma^1$ and $\Sigma^3$. $x^0_2(x^0_3)$ here is the coordinate
time function between $\Sigma^2$ and $\Sigma^3$.
Obviously, with the factor $A(x^0_3)$, it is a different coordinate time
function from Eq.(7), and it goes against the uniqueness of the
coordinate
time function between a pair of coordinate system. Therefore, we conclude
that is no associativity for the coordinate time function; it is a logical
flaw in the definition of proper time by the world line length.

A point implied in the above discussion is he correspondence of integral
domain of world line length
in different systems; it requires that range $[t_0,t], [t'_0,t']$ and
$[t''_0,t'']$ are exactly in one-to-one correspondence in the calculation
of the duration of proper time for the observed particle in these systems.
It is the only postulates we work on, and we call it {\bf Principle of
Measurement Correspondence}.

In terms of the proper time of the observors
this principle can be formulated as follows:
The duration of a particular physical process occupies a definite range of
the proper time axis of an observor, except for the choice of proper time
origin
and the ratio of ticking rate of clocks which is constant according
to the `hypothesis of consistency'[1]. Therefore, for any a couple of
different
observors, there is a unique function $\tau'=f(\tau)$ to relate the
measures of the proper times, $[\tau_1,\tau_2]$ and $[\tau'_1,\tau'_2]$,
they spend
for the observation of the process.

To illustrate the  physical implication of the principle, we set up an
imaginary
experiment which involves quantum measurement correlation (EPR effect). 
Suppose a
electron-positron couple with total spin zero is created but fails to form a
bound state in
an inertial system $\Sigma$ in Minkovski spacetime. Their classical
trajectories are therefore as follows:
$$x^e=vt,$$
\begin{eqnarray}
~~~~~~x^p=-vt,
\end{eqnarray}
if $\Sigma$
happen to be the mass center system. Let $\Sigma^1,\Sigma^2$ the co-moving
system of the electron and positron and the clocks attached to them are
set zero simultaneously with the clock in $\Sigma$ at the moment the
electron-positron couple is created. The wave functions of the particles in
the co-moving systems are the follows[2]:
\begin{eqnarray}
\psi^{e,(1)}_s(x')=\sqrt{2m}\left(
\begin{array}{c}
\chi_s  \\
0
\end{array}
\right)e^{-imt} \  \ (s=\pm {1\over 2}),
\end{eqnarray}
and
\begin{eqnarray}
\psi^{p,(2)}_s(x'')=-\sqrt{2m}\left(
\begin{array}{c}
0  \\
\varepsilon\chi_s
\end{array}
\right)e^{imt} \  \ (s=\pm {1\over 2}),
\end{eqnarray}
where $\chi_{1\over 2}=\left( \begin{array}{c} 1 \\0
\end{array}\right)$, $\chi_{-{1\over 2}}=\left( \begin{array}{c} 0 \\1
\end{array}\right)$, and $\varepsilon=\left(\begin{array}{lr}
0 & 1\\
-1 & 0
\end{array} \right)$. 

In system $\Sigma$, their wave functions, which satisfy Dirac equation in
a general system
\begin{eqnarray}
(i\gamma^{\mu}\frac{\partial}{\partial x^{\mu}}-m)\psi(x)=0,
\end{eqnarray}
can be obtained through Lorentz transformation:
\begin{eqnarray}
\psi_s^e(x)=S^{-1}(\Lambda(\beta))\psi^{e,(1)}_s(x'),
\end{eqnarray}
\begin{eqnarray}
\psi_s^p(x)=S^{-1}(\Lambda(-\beta))\psi^{p,(2)}_s(x''),
\end{eqnarray}
where $\Lambda^{\mu}_{\nu}=\delta^{\mu}_{\nu}+h^{\mu}_{\nu}$ is 
Lorentz transformation matrix of the coordinates
and $$S(\Lambda)=exp(-{i\over 4}h^{\mu\nu}\sigma_{\mu\nu})=exp({1\over
8}h^{\mu\nu}[\gamma_{\mu},\gamma_{\nu}]).$$
The total wave function of the spin-zero system in $\Sigma$ before any
measurement to determine the spin state of any of the particles is
\begin{eqnarray}
\Psi(x)={1\over \sqrt{2}}(\psi^e_{+{1\over 2}}\psi^p_{-{1\over 2}}-
\psi^e_{-{1\over 2}}\psi^p_{+{1\over 2}})(x).
\end{eqnarray}

At a particular moment $T_0$ in $\Sigma$, which corresponds to
$$T'_0=\gamma(T_0-{v^2\over c^2}T_0)=\sqrt{1-{v^2\over c^2}}~T_0,$$
in $\Sigma^1$, the observor in  $\Sigma^1$ undertake a measurement of the
spin state of the electron with the result, say $s=+{1\over 2}$, then
the total wave function will collapse to $\psi^{e,(1)}_{+{1\over
2}}(x')\psi^{p,(1)}_{-{1\over 2}}(x')$
immediately, because of quantum measurement correlation (EPR effect).
Therefore, in $\Sigma^1$, the total wave function since the creation the
electron-positron couple can be given as
$$\Psi^{(1)}(x')={1\over \sqrt{2}}(\theta(t')-\theta(t'-T'_0))
(\psi^e_{+{1\over 2}}\psi^p_{-{1\over 2}}-
\psi^e_{-{1\over 2}}\psi^p_{+{1\over 2}})(x')$$
\begin{eqnarray}
~~~~~~~~~~~~~~~+\theta(t'-T'_0)\psi^e_{+{1\over2}}
\psi^p_{-{1\over 2}}(x'),
\end{eqnarray}
where 
$\theta(t)=\left\{ \begin{array}{cc}
1, &    t>0, \\
0, &    t<0. 
\end{array}
\right.  $
In $\Sigma$, according to Lorentz transformation Eq.(15),
the correspondent wave function becomes
$$\Psi_1(x)=(\theta(\sqrt{1-{v^2\over c^2}}~t)-\theta(\sqrt{1-{v^2\over c^2}}~t-
\sqrt{1-{v^2\over c^2}}~T_0))\Psi(x)$$
\begin{eqnarray}
~~~~~~~~~+\theta(\sqrt{1-{v^2\over c^2}}~t-\sqrt{1-{v^2\over
c^2}}~T_0)S^{-1}(\Lambda(\beta))
\psi^{e,(1)}_{+{1\over 2}}\psi^{p,(1)}_{-{1\over 2}}(x').
\end{eqnarray}
Obviously, the moment of wave function collapse is determined by
$$\sqrt{1-{v^2\over c^2}}~t_1-\sqrt{1-{v^2\over c^2}}~T_0=0,$$
i.e. $t_1=T_0$. 

On the other hand, the coordinate time function between $\Sigma^1$ and
$\Sigma^2$ is given by
\begin{eqnarray}
t''=\frac{t'-(-{u\over c^2})(-ut')}{\sqrt{1-{u^2\over c^2}}}
=\frac{1-{v^2\over c^2}}{1+{v^2\over c^2}}~t',
\end{eqnarray}
since the relative velocity of $\Sigma^2$ to $\Sigma^1$ is
$u=\frac{-2v}{1+{v^2\over c^2}}$.
The total wave function in $\Sigma^2$ can be obtained by the Lorentz
transformation from $\Sigma^1$ to $\Sigma^2$ as  
$$\Psi^{(2)}(x'')=
((\theta(t'')-\theta(t''-\frac{1-{v^2\over c^2}}
{1+{v^2\over c^2}}~T'_0))S(\Lambda(-{u\over c}))
S(\Lambda({v\over c}))\Psi(x)$$
\begin{eqnarray}
+\theta(t''-\frac{1-{v^2\over c^2}}{1+{v^2\over c^2}}~T'_0))
S(\Lambda(-{u\over c}))
\psi^{e,(1)}_{+{1\over
2}}(x')\psi^{p,(1)}_{-{1\over 2}}(x').
\end{eqnarray}
With the consideration of the coordinate time function between
$\Sigma$ and $\Sigma^2$ 
\begin{eqnarray}
t''=\gamma (t-(-{v\over c})(-{v\over c})t),
\end{eqnarray}
the wave function $\Psi^{(2)}(x'')$ is transformed back to
$$\Psi_2(x)=(\theta(\sqrt{1-{v^2\over c^2}}~t)-
\theta(\sqrt{1-{v^2\over c^2}}~t
-\frac{(1-{v^2\over c^2})^{3\over 2}}{1+{v^2\over c^2}}~T_0))\Psi(x)$$ 
\begin{eqnarray}
+\theta(\sqrt{1-{v^2\over c^2}}~t
-\frac{(1-{v^2\over c^2})^{3\over 2}}{1+{v^2\over
c^2}}~T_0))S^{-1}(\Lambda(-{v\over
 c}))S(\Lambda(-{u\over c}))
\psi^{e,(1)}_{+{1\over
2}}(x')\psi^{p,(1)}_{-{1\over 2}}(x'),
\end{eqnarray}
which gives us another time for the collapse of wave function
at $t_2=\frac{1-{v^2\over c^2}}{1+{v^2\over c^2}}~T_0$ in $\Sigma$.

An absolute event (the operation of spin state measurement) in $\Sigma^1$ gives
rise to two different correspondences in $\Sigma$, and leads to
the ambiguity of the determination of proper time in $\Sigma^2$. 
In fact, it reflects the contradiction of the relativity of simultaneity 
with quantum measurement correlation effect. This phenomenon deserves
further
study on
the problem, if we desire to find a consistent theory.

\vspace{4mm}
\noindent
{\bf\large Appendix: Proof of the generalized glupico time.tex
ing lemma}
\renewcommand{\theequation}{A.\arabic{equation}}
\setcounter{equation}{0}

Gluing lemma[3] states that if there is $f: A\to B$ with
$$f(x)=\left\{ \begin{array}{cc}
f_1(x) &    x\in A_1, \\
f_2(x) &    x\in A_2, 
\end{array}
\right. $$ 
where $A=A_1\cup A_2$, the union of two open sets, and $f_1$ and $f_2$ are
continuous respectively, then
$f(x)$ is continuous in the whole open set $A$, as long as $f_1(x)=f_2(x)$ for 
$x\in A_1 \cap A_2$.

We need to generalize the statement to the case where the range $[t_0,t]$
is covered by a family of countably infinite open set plus two semi-open sets
at the end points:
$[t_0,t]=[t_0,t_1)\cup (t_1,t_2)\cdots \cup (t_{n-1},t_n)\cup (t_n,t]$, and
there is the implicit function $F(t,t')=0$ over $[t_0,t]\times [t'-0,t']$.
The open set $(t_i,t_{i+2})$ is so small that implicit function theorm
guarantees
a unique differentiable function $f_i(x)$ on it. From the uniqueness of
these functions we have $f_i(x)=f_{i+1}(x)$ for $x\in (t_{i+1},t_{i+2})$.
Let $C$ be an open set in $[t'_0,t']$ and $t$ the function which equals
$f_i$ on every these open set, we have
$$t^{-1}(C)\cap [t_0,t]=f^{-1}_0(C\cap [t_0,t_2))\cup f^{-1}_1(C\cap (t_1,t_3))
\cdots \cup f^{-1}_{n-1}(C\cap (t_{n-1},t]).$$
It is an open set because the countably infinite union of open sets is an
open set, hence the continuity of function $t$.

\vspace {4mm}
{\bf ACKNOWLEDGMENTS}. The authors thank Q.B.Li for constructive
discussions, and S. H. Dong, G.S.Huang, A.L.Zhang for extensive helps in 
completing the paper.

\end{document}